\documentclass{emulateapj}

\usepackage{natbib,graphicx,textcomp,lscape,amsmath,amsthm,ulem, subfigure,color}

\newcommand{\logr}{log(\ensuremath{R'_{\mbox{\scriptsize HK}}})}

\begin{document}

\title{\textit{Hubble Space Telescope} Near-IR Transmission Spectroscopy of the Super-Earth HD 97658b} 

\author{
Heather A. Knutson\altaffilmark{1,2}, Diana Dragomir\altaffilmark{3,4}, Laura Kreidberg\altaffilmark{5}, Eliza M.-R. Kempton\altaffilmark{6}, P. R. McCullough\altaffilmark{7}, Jonathan J. Fortney\altaffilmark{8}, Jacob L. Bean\altaffilmark{5}, Michael Gillon\altaffilmark{9}, Derek Homeier\altaffilmark{10}, Andrew W. Howard\altaffilmark{11}
} 

\altaffiltext{1}{Division of Geological and Planetary Sciences, California Institute of Technology, Pasadena, CA 91125, USA} 
\altaffiltext{2}{hknutson@caltech.edu}
\altaffiltext{3}{Las Cumbres Observatory Global Telescope Network, Goleta, CA 93117, USA}
\altaffiltext{4}{Department of Physics, Broida Hall, UC Santa Barbara, CA, USA}
\altaffiltext{5}{Department of Astronomy and Astrophysics, University of Chicago, Chicago, IL 60637, USA}
\altaffiltext{6}{Department of Physics, Grinnell College, Grinnell, IA 50112 USA}
\altaffiltext{7}{Space Telescope Science Institute, Baltimore, MD 21218, USA}
\altaffiltext{8}{Department of Astronomy and Astrophysics, University of California, Santa Cruz, CA 95064 USA}
\altaffiltext{9}{Institut d'Astrophysique et de G\' eophysique, Universite\' e de Li\' ege, Li\' ege 1, Belgium}
\altaffiltext{10}{Centre de Recherche Astrophysique de Lyon, 69364 Lyon, France}
\altaffiltext{11}{Institute for Astronomy, University of Hawaii at Manoa, Honolulu, HI, USA}

\begin{abstract}

Recent results from the \textit{Kepler} mission indicate that super-Earths (planets with masses between $1-10$ times that of the Earth) are the most common kind of planet around nearby Sun-like stars.  These planets have no direct solar system analogue, and are currently one of the least well-understood classes of extrasolar planets.  Many super-Earths have average densities that are consistent with a broad range of bulk compositions, including both water-dominated worlds and rocky planets covered by a thick hydrogen and helium atmosphere.  Measurements of the transmission spectra of these planets offer the opportunity to resolve this degeneracy by directly constraining the scale heights and corresponding mean molecular weights of their atmospheres.  We present \textit{Hubble Space Telescope} near-infrared spectroscopy of two transits of the newly discovered transiting super-Earth HD 97658b.  We use the Wide Field Camera 3's scanning mode to measure the wavelength-dependent transit depth in thirty individual bandpasses.  Our averaged differential transmission spectrum has a median $1\sigma$ uncertainty of 23 ppm in individual bins, making this the most precise observation of an exoplanetary transmission spectrum obtained with WFC3 to date.  Our data are inconsistent with a cloud-free solar metallicity atmosphere at the $10\sigma$ level.  They are consistent at the $0.4\sigma$ level with a flat line model, as well as effectively flat models corresponding to a metal-rich atmosphere or a solar metallicity atmosphere with a cloud or haze layer located at pressures of 10 mbar or higher.  
\end{abstract}

\keywords{binaries: eclipsing --- planetary systems --- techniques: spectroscopy}

\section{Introduction}\label{intro}

The \textit{Kepler} mission has resulted in the discovery of more than three thousand transiting planets and planet candidates \citep{batalha13,burke14} to date, with a majority of the sample consisting of sub-Neptune-sized planets.  Analyses of this set of Kepler planet candidates indicate that planets with radii intermediate between that of Neptune and the Earth appear to be the most common kind of extrasolar planet orbiting nearby FGK stars, with a peak at radii between $2-3\times$ that of the Earth \citep{howard12,fressin13,petigura13}.  Planets in this size range are typically referred to as ``super-Earths",  although they could potentially form with a broad range of compositions including primarily rocky with a thin atmosphere (true ``super-Earths"), a rocky or icy core surrounded by a thick hydrogen atmosphere (``mini-Neptunes"), or water-dominated with a thick steam atmosphere (``water worlds").  Many of these super-Earths are found in close-in, tightly packed multiple planet systems \citep[e.g.,][]{lissauer11,fabrycky12,steffen13}, and there is an ongoing debate as to whether these systems formed in place or migrated in from more distant orbits \citep{hansen12,chiang13,raymond14}.

Detailed studies of super-Earth compositions offer important clues on their origins: presumably water-rich planets must have formed beyond the ice line, while in-situ formation models predict primarily rocky compositions with relatively water-poor hydrogen-dominated atmospheres \citep[e.g.,][]{raymond08}.  By combining radius measurements from \textit{Kepler} with mass estimates obtained using either the radial velocity or transit timing techniques, it is possible to constrain the average densities and corresponding bulk compositions of the super-Earths in the Kepler sample \citep[e.g.,][]{lithwick12,hadden13,weiss14,marcy14}.  These observations indicate that the super-Earths in the Kepler sample display a broad range of average densities, with a transition towards denser, primarily rocky compositions below $1.5-2$ Earth radii \citep{weiss14,marcy14}.  For the larger, lower-density super-Earths in the Kepler sample it is possible to match their measured densities with a broad range of compositions, including both water-rich and water-poor scenarios, simply by varying the amount of hydrogen in their atmospheres \citep{seager07,valencia07,valencia13,rogers10a,zeng13}.  Although it can be argued that some compositions are unlikely based on models of planet formation and atmospheric mass loss for close-in planets, this still leaves a wide range of plausible models \citep[e.g.,][]{rogers10b,rogers11,nettelmann11,heng12,lopez12,lopez13,fortney13}.  

Measurements of the transmission spectra of super-Earths allow us to directly estimate the mean molecular weight of the planet's atmosphere \citep[e.g.,][]{miller08,fortney13}, which in turn provides improved constraints on its interior composition.  Planets with cloud-free, hydrogen dominated atmospheres will have relatively large scale heights and correspondingly strong absorption features during transit, while planets with hydrogen-poor atmospheres will have relatively small scale heights and weak absorption.  Unfortunately, the majority of the super-Earths detected by Kepler orbit faint stars ($m_K>10$), making it difficult to accurately measure their transmission spectra using existing facilities.  There are currently three super-Earths known to transit relatively bright stars, including:  GJ 1214b \citep[$m_K=8.8$;][]{charbonneau09}, 55 Cancri e \citep[$m_K=4.0$;][]{winn11,demory11}, and HD 97658b \citep[$m_K=5.7$;][]{dragomir13}.  

GJ 1214b is currently the only one of these three systems with a well-characterized transmission spectrum, which is flat and featureless across a broad range of wavelengths \citep[e.g.,][]{bean10,bean11,desert11,berta12,kreidberg14}.  Although initial observations were consistent with either a cloud-free, hydrogen-poor atmosphere or a hydrogen-rich atmosphere with a high cloud deck \citep[e.g.,][]{bean10}, the most recent near-IR observations obtained by \citet{kreidberg14} using the Wide Field Camera 3 (WFC3) instrument on the \textit{Hubble Space Telescope (HST)} are precise enough to rule out cloud-free models over a broad range of atmospheric metallicities \citep[also see][]{benneke13}.  The presence of a high altitude cloud or haze layer means that for this planet, at least, transmission spectroscopy provides relatively weak constraints on the mean molecular weight of its atmosphere and, by extension, on its interior composition.

\begin{figure}[ht]
\epsscale{1.2}
\plotone{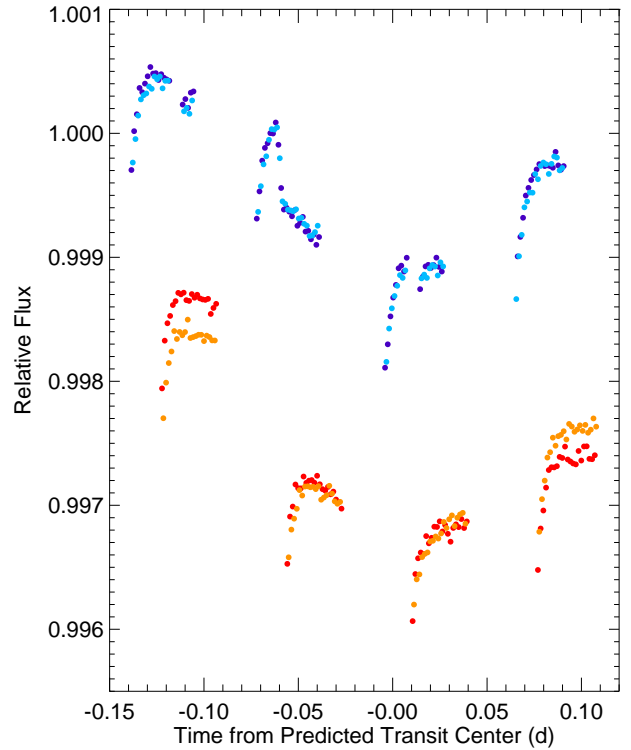}
\centering
\caption{Raw white light photometry for the UT 2013 Dec 19 visit (top) and UT 2014 Jan 7 transit (bottom).  Different scan directions for the December visit are indicated as light (forward scan) and dark (reverse scan) blue filled circles.  Scan directions for the January visit are plotted as yellow (forward scan) and red (reverse scan) filled circles.  The two light curves have been offset by a relative flux of 0.0015 for clarity.  The first spacecraft orbit has been trimmed from each visit, leaving four orbits per transit.  We have also normalized the light curves from each scan direction to one by dividing each by the median flux value; we do this in order to remove a small offset in the measured fluxes from the two scan directions.}
\label{raw_phot}
\end{figure} 

In this paper we present \textit{HST} WFC3 near-infrared transmission spectroscopy of the transiting super-Earth HD 97658b.  This planet was first detected using the radial velocity technique \citep{howard11}, and later found to transit using \textit{MOST} photometry \citep{dragomir13} .  It has a mass of $7.9\pm0.7$ ~$M_\earth$, a radius of $2.3\pm0.2$~$R_\earth$, and an average density of $3.4\pm0.9$ g cm$^{-1}$ \citep{dragomir13}, making it modestly denser and more massive than GJ 1214b.  HD 97658b orbits its K star primary with a period of 9.5 days, and has a predicted zero-albedo temperature between $700-1000$~K depending on the efficiency of energy transport to the planet's night side.  If this planet has the same atmospheric composition as GJ 1214b, its modestly higher atmospheric temperature might prevent the formation of the cloud layer detected in GJ 1214b's atmosphere  \citep{morley13}.  Our observations are obtained at wavelengths between $1.2-1.6$~\micron, and are primarily sensitive to the presence or absence of the water absorption band located at 1.4~\micron; this feature has now been robustly detected in the atmospheres of several hot Jupiters \citep[e.g.,][]{deming13,wakeford13,mandell13,mccullough14}, and is expected to be present in super-Earths as well over a broad range of atmosphere compositions \citep[e.g.,][]{benneke12,kreidberg14,hu14}.  We discuss our observations in \S\ref{obs} and the implications for the properties of this planet's atmosphere in \S\ref{discussion}.

\begin{figure}[ht]
\epsscale{1.2}
\plotone{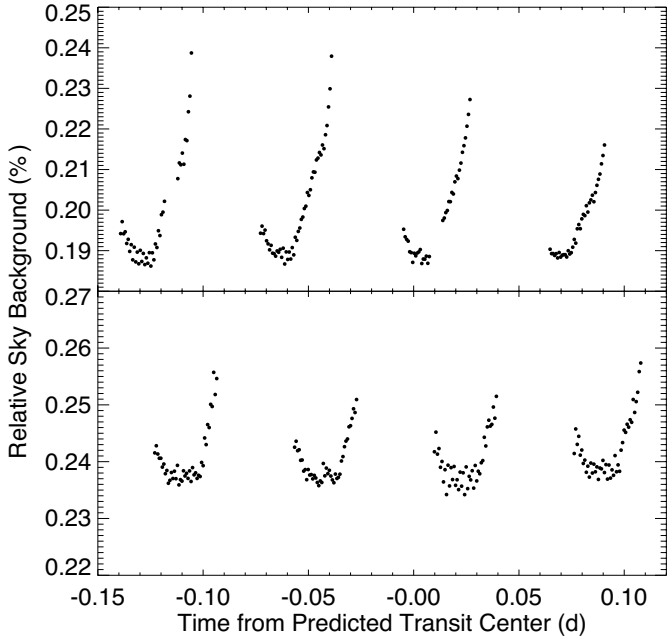}
\centering
\caption{Estimated sky background as a percentage of the total flux in the white-light aperture (i.e., summed over all wavelengths) for stacked images created using the method described in section \ref{deming_method}.  Sky backgrounds for the December visit are shown in the top panel and background for the January visit are shown in the bottom panel.}
\label{sky_bkd}
\end{figure}

\section{Observations}\label{obs}

We observed one transit of HD 97658b on UT 2013 Dec 19 and another on UT 2014 Jan 7 (GO 130501, PI Knutson) using the G141 grism on the \textit{HST} WFC3 instrument, which provides low resolution spectroscopy at wavelengths between $1.1-1.7$~\micron.  Each observation consists of five \textit{HST} orbits with a total duration of approximately seven hours per visit; our target was visible for approximately half of each 96 minute \textit{HST} orbit.  Scheduling constraints resulted in a slightly shorter first orbit during the January visit, which consisted of 203 spectra instead of the 206 spectra obtained during the December visit.  During the January visit we also utilized a series of short exposures at the end of each orbit in order to force a buffer dump, which avoided the mid-orbit gaps in coverage visible in the light curves for the first visit in Fig. \ref{raw_phot}.  We do not include these short exposures in our analysis, as they were taken in imaging rather than spectroscopic mode.  

Our spectra were obtained using the $256\times256$ pixel subarray and SPARS10 mode with four samples, giving a total integration time of 14.97~s in each image.  Our target is one of the brightest transiting planet host stars, and we therefore utilized the new scanning mode \citep[e.g.,][]{mccullough12,deming13,kreidberg14,knutson14} with a scan rate of $1.4\arcsec$ s$^{-1}$ in order to achieve a higher observing efficiency while remaining well below saturation.  This results in a scanned spectral image that fills most of the subarray image, with a fractional coverage similar to that of the HD 209458 observations from \citet{deming13}.  We also alternated between forward and reverse scan directions in order to further reduce overheads; this approach was previously used by \citet{kreidberg14} for GJ 1214b.  The resulting spectra have peak counts around 40,000 electrons pixel$^{-1}$, which is high enough to cause a modestly steep asymptotic ramp in the measured fluxes within individual orbits \citep[see ][for a discussion of the relationship between the ramp slope and total flux]{wilkins14}. 

We reduced the data using two independent methods, which allow us to test whether or not the transmission spectrum we obtain from our analysis is sensitive to our choice of analysis technique.  We discuss each approach separately below.

\begin{figure}[ht]
\epsscale{1.2}
\plotone{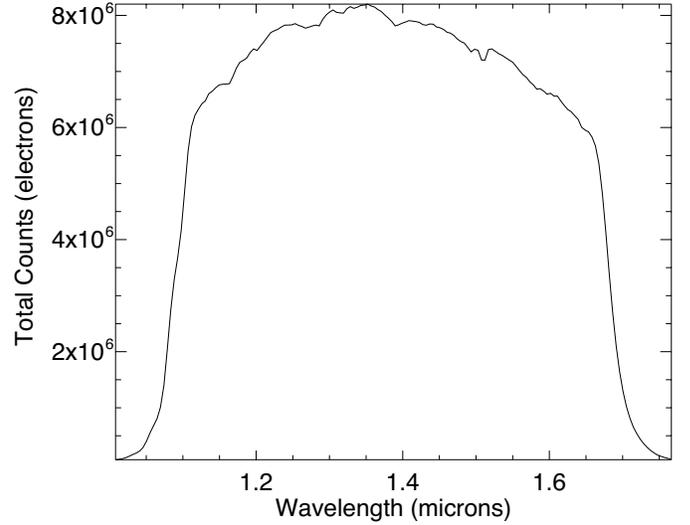}
\centering
\caption{A representative spectrum from the UT 2013 Dec 19 visit; this spectrum was created from the two dimensional image by summing in the $y$ (cross-dispersion) direction.}
\label{sample_spec}
\end{figure}

\subsection{Spectral Template Fitting Method}\label{deming_method}

In this approach we utilized the spectral template method first presented in \citet{deming13}, which we describe in detail in \citet{knutson14} and summarize here.  We begin with the raw sample up the ramp images from the ima.fits files, which are bias and dark subtracted, and apply our own flat-fielding and wavelength calibrations based on the standard WFC3 pipeline as discussed in \citet{wilkins14}.  We then subtract successive non-destructive readout pairs in order to create a series of difference images, trim the region around the spectral scan in each image, and sum the resulting images to create a composite containing the full scan.  The benefit of this approach is that it minimizes the contributions of pixels that are not directly illuminated by the star in a given readout time step \citep[see][for more discussion on this point]{deming13}.  We estimate that the sky background in our final composite images is $0.18-0.26\%$ of the total flux measured in each wavelength element (see Fig. \ref{sky_bkd}), and subtract this estimated sky background from each image.

\begin{figure}[ht]
\epsscale{1.2}
\plotone{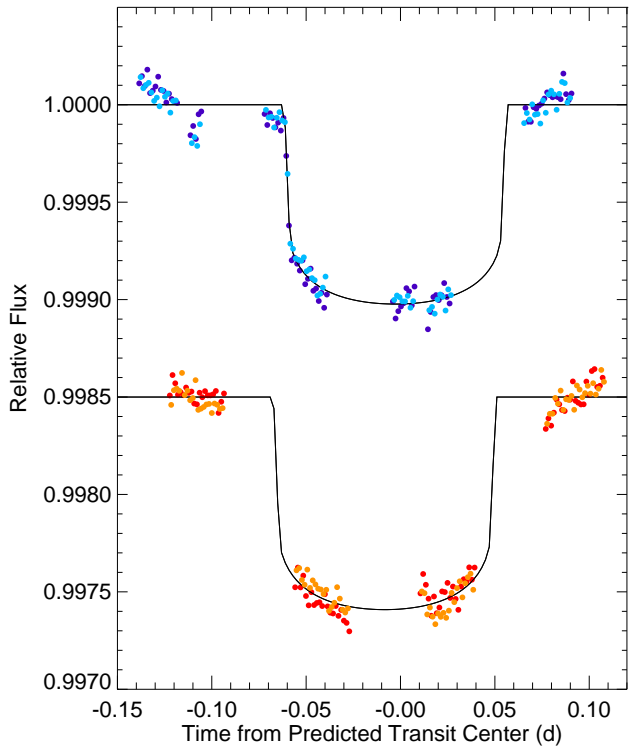}
\centering
\caption{Normalized white light photometry with best-fit detector effects removed for the UT 2013 Dec 19 visit (top) and UT 2014 Jan 7 transit (bottom).  Different scan directions for the December visit are indicated as light (forward scan) and dark (reverse scan) blue filled circles.  Scan directions for the January visit are plotted as yellow (forward scan) and red (reverse scan) filled circles.  The first spacecraft orbit has been trimmed from each visit, leaving four orbits per transit, and the two light curves have been offset by a relative flux of 0.0015 for clarity.}
\label{norm_phot}
\end{figure}

Once we have created a stacked image from each set of readouts, we apply a filter to correct for bad pixels and cosmic ray hits as described in \citet{knutson14}.  We then sum the spectrum along the $y$ axis in order to extract a one dimensional spectrum, using an aperture that extends fifteen pixels above and below the edges of the spectrum (200 pixels in total) in order to include the extended wings of the point spread function (see Fig. \ref{sample_spec} for a representative example).  Because the WFC3 spectra are undersampled \citep{deming13}, we convolve each of our one dimensional spectra with a Gaussian function with a width (FWHM) of four pixels in order to mitigate effects related to the shifting position of the spectrum on the detector.  Although this modestly degrades the spectral resolution of our data, we later bin our transmission spectrum by an equivalent amount in order to increase the signal to noise ratio.  We calculate the MJD mid-exposure time corresponding to each image using the information from the flt.fits image headers, and convert these times to the BJD$_{TDB}$ time standard following the methods of \citet{eastman10}.

Our next step is to create a spectral template by averaging the ten spectra immediately before and after the transit.  We then fit this template spectrum to each individual spectrum in our time series, allowing the relative position and amplitude of the template spectrum to vary as free parameters.  We find no difference in our results if we create a separate spectral template for the forward and reverse scan images, and we therefore use the same template for all images.  The resulting series of best-fit amplitudes are plotted in Fig. \ref{raw_phot}, and are identical to the white-light curves obtained by summing the fluxes across all wavelengths.  We find that there is a small flux offset between the light curves for the two scan directions, which we remove by dividing each time series by its median flux value.  \citet{mccullough12} suggest that this offset is a consequence of the order in which columns are read out by the detector (the ``up-stream/down-stream effect"); when the scan moves in the same direction as the readout then the effective integration time will be slightly longer than in the case of a reverse scan.  The forward and reverse spectra in our images also occupy slightly different positions on the array, which might also contribute to this offset.  

We subtract the best-fit spectral template from each individual spectrum in order to create a differential time series for each individual wavelength element.  The benefit to this approach is that it effectively removes all common-mode detector effects from the differential light curve.  We find that the scatter in our light curves for individual wavelength elements is within 5\% of the photon noise limit in all cases, indicating that there is minimal color-dependence in the systematic noise sources.

\begin{figure}[ht]
\epsscale{1.15}
\plotone{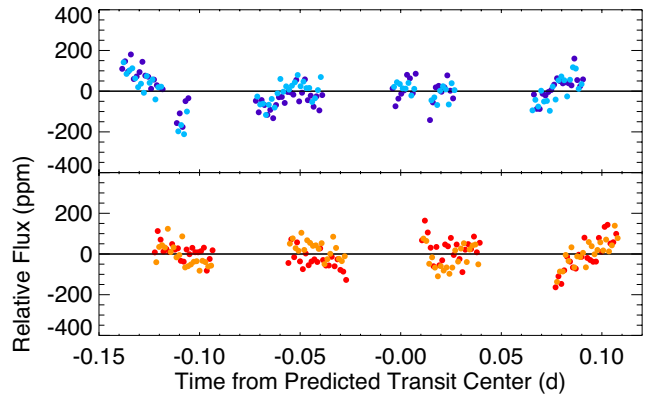}
\centering
\caption{White light residuals after the best-fit detector and transit light curves are removed for the UT 2013 Dec 19 visit (top panel) and UT 2014 Jan 7 transit (bottom panel).  Different scan directions for the December visit are indicated as light (forward scan) and dark (reverse scan) blue filled circles.  Scan directions for the January visit are plotted as yellow (forward scan) and red (reverse scan) filled circles.}
\label{residuals}
\end{figure} 

\subsubsection{White Light Transit Fits}\label{white_light_fits}

We take the two raw transit light curves plotted in Fig. \ref{raw_phot} and fit them each with a transit light curve, a linear function of time, and an exponential ramp in orbital phase: 

\begin{equation}
F(t) = c_1(1+ c_2t+c_3e^{-p/c_4})F_{transit}(t)
\end{equation} 
where $c_1-c_4$ are free parameters in the fit, $t$ is the time from the measured transit center in days and $p$ is the time in days from either the first observation in each spacecraft orbit or the last observation before a mid-orbit buffer dump.  We use this definition because we find that the ramp after a mid-orbit buffer dump is usually not as steep as the initial ramp at the start of the orbit, and using this definition provided a good fit to the observed behavior by reducing the amplitude of the mid-orbit ramp model.  The latter definition is only used for the December transit observation, which includes two mid-orbit buffer dumps that cause the ramp effect to reset.  We allow the planet-star radius ratio $R_p/R_\star$ and center of transit time to vary as free parameters for each individual transit.  We find that the behavior of the instrumental systematics is slightly different for the white light curves calculated from the forward and reverse scan directions (see Fig. \ref{raw_phot}).   We therefore allow the linear function of time and exponential ramp functions to vary independently for each scan direction, while keeping the same $R_p/R_\star$ and center of transit time. 

We calculate the transit light curve $F_{transit}$ following \citet{mandel02} where we initially fixed the orbital inclination $i$ and the ratio of the semi-major axis to the stellar radius $a/R_\star$ equal to their best-fit values from \citet{vangrootel14}.  We found that the uncertainty on our reported transit time for the first visit, which spanned the transit ingress, was particularly sensitive to our choice of values for these parameters.  Our transit timing uncertainties for this visit ranged between 30~s up to 5 minutes for the cases where $i$ and $a/R_\star$ were either fixed or allowed to vary freely in our fits.  We addressed this by acquiring the normalized 4.5~\micron~\textit{Spitzer} transit photometry from \citet{vangrootel14} and fitting this light curve simultaneously with our white-light \textit{HST} photometry assuming a single global value for $i$ and for $a/R_\star$.  This ensured that our final reported transit times correctly accounted for the added uncertainties contributed by these two parameters.  We allowed the \textit{Spitzer} planet-star radius ratio and center of transit time to vary independently in our fits, and found that our results for these two parameters were consistent with those reported by \citet{vangrootel14}.  In our final fits the uncertainty on the transit time for the first visit increased from 30~s to one minute as compared to the fixed $i$ and $a/R_\star$ case and the best-fit transit time was 3.8 minutes earlier.  The second visit does not include observations during ingress or egress, and has a correspondingly large uncertainty on the best-fit transit time.  We find that for this visit, allowing $i$ and $a/R_\star$ to vary has a negligible effect on the reported uncertainties in the best-fit transit time.

We calculate our initial predicted transit times for the \textit{HST} observations using the ephemeris from \citet{vangrootel14}, and then derive an updated estimate for the orbital period and center of transit time based on our new observations.  We then repeat our transit fits using this updated orbital period.  We calculate four-parameter nonlinear limb-darkening parameters \citep{claret00} for our transit light curves, with the relative intensity at each position given as:
\begin{equation}
\frac{I(\mu)}{I(1)} = 1-\sum_{k=1}^{4}a_k(1-\mu^{k/2})
\end{equation}
where $I(1)$ is the specific intensity at the center of the stellar disk, $a_k$ is the $k$th limb-darkening coefficient, and $\mu=cos(\theta)$ where $\theta$ is the angle between the the line of sight and the location of the emerging flux.  We calculate our limb-darkening coefficients using a \texttt{PHOENIX} stellar atmosphere model\citep{allard12}, where we take the flux-weighted average of the theoretical stellar intensity profile across the band and then fit for the limb-darkening coefficients.  We use the best-fit stellar parameters from \citet{vangrootel14}, who find an effective temperature of $5170\pm50$~K, a surface gravity of $4.58\pm0.05$, and a metallicity of $-0.23\pm0.03$.  These values are also consistent with a recent analysis by \citet{mortier13}, although this study prefers a lower metallicity of $-0.35\pm0.02$ and does not take into account the constraints on the stellar density from the transit light curve.  We also tried models with effective temperatures ranging between $5120-5220$~K and a metallicity of $-0.35$, but found that these had a negligible effect (2 ppm or less) on our resulting transmission spectrum.  The inclusion of limb-darkening in our fits creates a small offset in the average transit depth as compared to fits without limb-darkening, and has a negligible effect on the slope of the resulting transmission spectrum across the bandpass.  We find that the uncertainty in the stellar effective temperature contributes a systematic error of 1 ppm to our estimate of the differential transit depths.

\begin{deluxetable}{lrrrrcrrrrr}
\tabletypesize{\scriptsize}
\tablecaption{Transit Parameters from Joint \textit{HST} and \textit{Spitzer} Fits \label{transit_fit}}
\tablewidth{0pt}
\tablehead{
\colhead{Parameter} & \colhead{Value}}
\startdata
\textit{Global Values} & \\
$i$(\degr) & $89.85\pm0.48$\tablenotemark{a} \\
$a/R_{\star}$ & $26.24\pm1.21$ \\
$P$ (days)\tablenotemark{b} & $9.489264\pm0.000064$\\
$T_0$ (BJD$_{TDB}$)\tablenotemark{b} & $2456665.46415\pm0.00078$\\
&\\
\textit{UT 2013 Aug 10 \textit{Spitzer} Transit}\tablenotemark{c} & \\
$R_p/R_{\star}$ & $0.02778\pm0.00073$ \\ 
$T_c$ (BJD$_{TDB}$) & $2456523.12537\pm0.00049$\\
&\\
\textit{UT 2013 Dec 19 \textit{HST} Transit} & \\
$R_p/R_{\star}$ & $0.03012\pm0.00087$ \\ 
$T_c$ (BJD$_{TDB}$) & $2456646.48556\pm0.00071$\\
&\\
\textit{UT 2014 Jan 7 \textit{HST} Transit} & \\
$R_p/R_{\star}$ & $0.03111\pm0.00080$ \\ 
$T_c$ (BJD$_{TDB}$) & $2456665.4584\pm0.0056$\\
\enddata
\tablenotetext{a}{We limited the value of the inclination to be less than or equal to 90$\degr$ in our fits.  We find a $1\sigma$ lower limit on the inclination of 89.65\degr, and a $2\sigma$ limit of 88.79\degr.}
\tablenotetext{b}{We calculate our updated ephemeris using the published transit times from \citet{dragomir13} and \citet{vangrootel14} and our two new transit measurements.  $T_0$ is the zero epoch transit center time from our best-fit ephemeris, and $T_c$ is the measured transit center time from a given observation.}
\tablenotetext{c}{We find that for the \textit{Spitzer} data the uncertainties from the covariance matrix are larger than those from the prayer bead analysis, and report the covariance errors here.}
\end{deluxetable}

We estimate the uncertainties in our fitted white light parameters using a residual permutation (``prayer bead") method, the covariance matrix from our Levenberg-Marquart minimization, and a Markov Chain Monte Carlo (MCMC) analysis with $10^6$ steps.  In our initial fits with fixed $i$ and $a/R_\star$, we set the uncertainties on individual points in our \textit{HST} white light time series equal to the standard deviation of the residuals from our best-fit solution.  We find that the residual permutation technique results in uncertainties that are $2-3$ times larger than the corresponding values from both the covariance matrix and MCMC analysis.  This is expected, as the noise in our white light curves is dominated by time-correlated instrument effects, while the MCMC and covariance methods implicitly assume random Gaussian-distributed noise \citep{carter09}.  We plot the 2D probability distributions from the MCMC analysis and confirm that there are no significant correlations between any of the fit parameters in this case.  For the case where we fit the \textit{Spitzer} and \textit{HST} transits simultaneously while allowing $i$ and $a/R_\star$ to vary as free parameters, we increase the per-point uncertainties on the \textit{HST} transit light curves by a factor of two in order to more accurately reflect the uncertainties contributed by the time-correlated component of the noise.  We selected this scaling factor by requiring that the uncertainties on $R_p/R_\star$ derived from the covariance matrix be equal to the uncertainties from the prayer bead method.  This step is critical for the joint fits, because otherwise the \textit{HST} data have a disproportionate influence on the preferred values of $i$ and $a/R_\star$.  We find that in this version of the fit the values of the inclination and a/R* show a high degree of correlation, which is a well known property of this particular parameterization of the transit light curve shape.  None of the transit parameters in either version of the fits are correlated with the parameters used to describe the time-varying instrumental signal.  

\begin{figure}[ht]
\epsscale{1.2}
\plotone{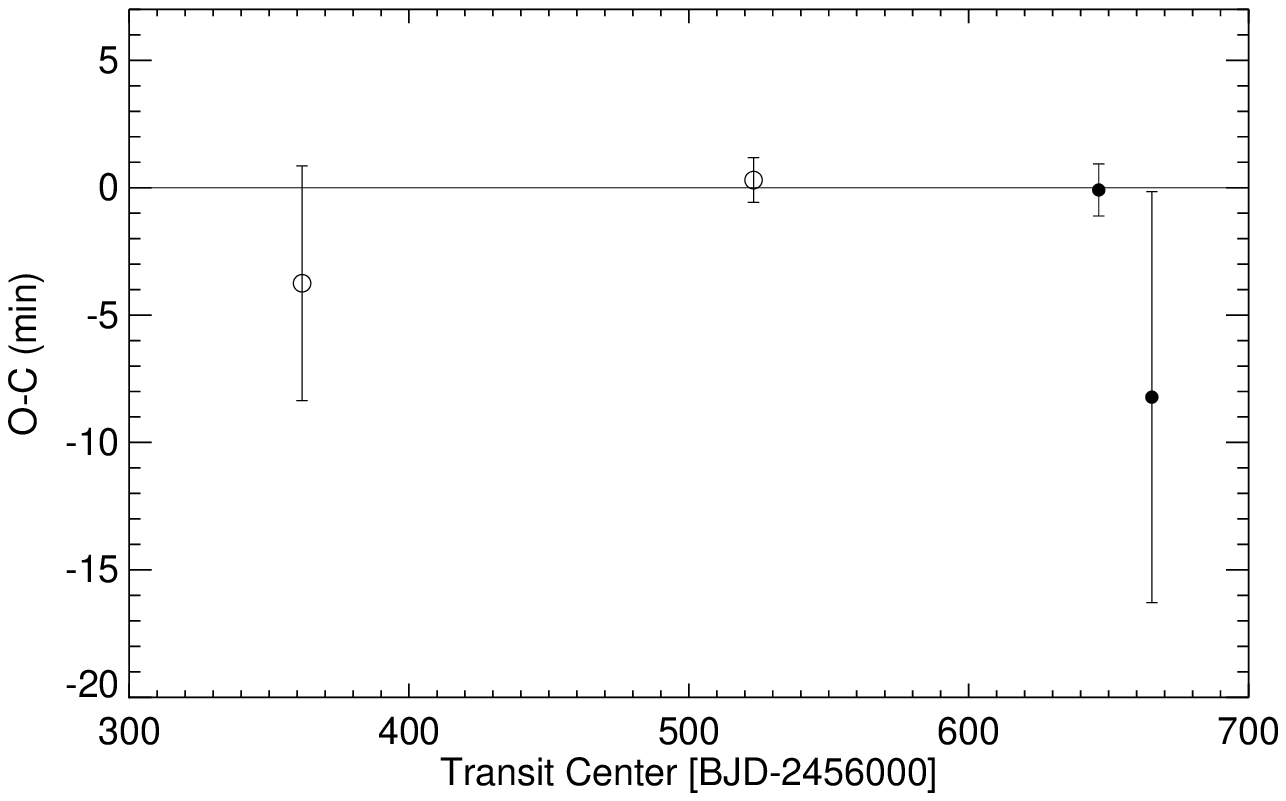}
\centering
\caption{Observed minus calculated transit times using our updated ephemeris.  Previously published \textit{MOST} \citep{dragomir13} and \textit{Spitzer} transit times \citep{vangrootel14} are shown as open circles, with our new \textit{HST} observations are shown as filled circles.  The second \textit{HST} transit has a larger uncertainty on the transit time because it does not include any points during ingress or egress, which provide the strongest constraints on the measured center of transit time.}
\label{ominusc}
\end{figure}

We give our best-fit transit parameters for the joint \textit{HST} and \textit{Spitzer} fits in Table \ref{transit_fit}, and plot the normalized transit light curves after dividing out our best fit for detector effects in Fig. \ref{norm_phot}.  The white light curve residuals from UT 2013 Dec 19 and UT 2014 Jan 7 have a rms of [73,68] ppm and [63,57] ppm, respectively, where we calculate this number separately for data from the forward and reverse scan directions.  We show the residuals from these fits in Fig. \ref{residuals}, and the offset between the observed and calculated transit times from our updated orbital ephemeris in Fig. \ref{ominusc}.  %The second transit does not have any measurements during ingress or egress, and therefore has a correspondingly large uncertainty on the measured center of transit time.

\begin{deluxetable}{ccccc}
\tablecaption{Binned Four Parameter Nonlinear Limb-Darkening Coefficients\tablenotemark{a}\label{ld_table}}
\tablehead{
\colhead{Wavelength\tablenotemark{b}} & \colhead{$c_1$} & \colhead{$c_2$}  & \colhead{$c_3$} & \colhead{$c_4$}
}
\startdata
       1.133 &       0.6247 &      -0.4071&       0.6890 &      -0.3099 \\
       1.152 &       0.6489 &      -0.4619 &       0.7545 &      -0.3395 \\
       1.171 &       0.6710 &      -0.5019 &       0.7856 &      -0.3530 \\
       1.190 &       0.6925 &      -0.5644 &       0.8445 &      -0.3743 \\
       1.208 &       0.7275 &      -0.6442 &       0.9250 &      -0.4065 \\
       1.227 &       0.7631 &      -0.7246 &        1.0242 &      -0.4507 \\
       1.246 &       0.8020 &      -0.8191 &        1.1222 &      -0.4894 \\
       1.265 &       0.8576 &      -0.9453 &        1.2552 &      -0.5437 \\
       1.284 &       0.4703 &       0.1247 &      0.0695 &     -0.0947 \\
       1.303 &       0.5182 &     -0.01457 &       0.2336 &      -0.1563 \\
       1.321 &       0.5640 &      -0.1255 &       0.3410 &      -0.1967 \\
       1.340 &       0.6366 &      -0.2849 &       0.5109 &      -0.2661 \\
       1.359 &       0.7125 &      -0.4597 &       0.6892 &      -0.3356 \\
       1.378 &       0.8012 &      -0.6639 &       0.8937 &      -0.4138 \\
       1.397 &       0.8842 &      -0.8546 &        1.0907 &      -0.4911 \\
       1.416 &       0.8901 &      -0.8644 &        1.1018 &      -0.4975 \\
       1.434\tablenotemark{c} &        1.0237 &       -1.1944 &        1.4386 &      -0.6244 \\
       1.453 &        1.1298 &       -1.4436 &        1.6910 &      -0.7202 \\
       1.472 &        1.2482 &       -1.7152 &        1.9657 &      -0.8248 \\
       1.491\tablenotemark{c} &       0.4779 &       0.4024 &      -0.4201 &       0.1113 \\
       1.510 &       0.5016 &       0.3492 &      -0.3886 &       0.1055 \\
       1.529 &       0.5557 &       0.3381 &      -0.4372 &       0.1307 \\
       1.547 &       0.6129 &       0.2520 &      -0.3962 &       0.1249 \\
       1.566 &       0.6479 &       0.1523 &      -0.3120 &      0.1000 \\
       1.585 &       0.6325 &       0.1655 &      -0.3350 &       0.1121 \\
       1.604 &       0.7095 &      0.0517 &      -0.2727 &      0.0993 \\
       1.623 &       0.7208 &      0.0239 &      -0.2472 &      0.0905 \\
       1.642 &       0.7129 &      0.0421 &      -0.2933 &       0.1144 \\
       &&&&\\
       4.5\tablenotemark{d} & 0.8713 & -1.4634 & 1.4955 & -0.5581\\
\enddata
\tablenotetext{a}{The spectra extracted from the UT 2013 Dec 19 and UT 2014 Jan 7 visits have slightly different wavelength solutions, and we therefore created custom limb darkening coefficients for each visit.  A full table of both the binned and unbinned coefficients for both visits is available upon request.}
\tablenotetext{b}{These binned coefficients are only used in the white light residual fitting analysis.  In the spectral template fitting analysis we fit the light curves for individual wavelength elements and use limb-darkening coefficients calculated appropriately for this resolution (4x higher than shown here).} 
\tablenotetext{c}{There is a well-known degeneracy in the values of the four-parameter nonlinear limb-darkening coefficients; although the values of the coefficients do not vary smoothly, we find that the corresponding stellar intensity profiles are continuous across all wavelengths considered here.}
\tablenotetext{d}{Limb-darkening coefficients calculated for the 4.5~\micron~\textit{Spitzer} IRAC bandpass, which we use to fit the transit light curve from \citet{vangrootel14}.}
\end{deluxetable}

\subsubsection{Differential Transit Fits}

We estimate the planet's wavelength-dependent transit depth by fitting the differential time series for each individual wavelength element.  In this case we fit the differential time series for all five orbits rather than the four used in the white light fits, as we find that the differential light curves do not show any detectible systematic trends during the first orbit.  The light curves for the forward and reverse scans are offset by a constant flux value, so we divide each light curve by its median value before carrying out our fit.  We fit each light curve with a linear function of time and a transit function calculated as the difference between the transit light curve in that band and the white-light transit curve, where we fix the values of $i$ and $a/R_\star$ to their best-fit values from the white-light analysis.  We also tried fits where we allowed an independent linear function of time for each scan direction, but found that this gave a transmission spectrum that was indistinguishable from the case where we assumed the same linear function for both scan directions.  We use the best-fit transit time from the white light fits, and fit for the planet-star radius ratio $R_p/R_\star$ corresponding to each wavelength element.  The best-fit transit depths reported here are simply the square of these values.  

We calculate the appropriate four-parameter nonlinear limb-darkening coefficients for each wavelength element using the same PHOENIX model used for the white-light fits (see Table \ref{ld_table}), where we convolve the model spectrum at each position on the star with the same Gaussian function used on our data before fitting for the limb-darkening coefficients.  We find that varying the stellar effective temperature by $\pm50$~K changes the resulting transit depths by 1 ppm, which we include in our analysis as a separate systematic error term. 

\begin{figure}[ht]
\epsscale{1.2}
\plotone{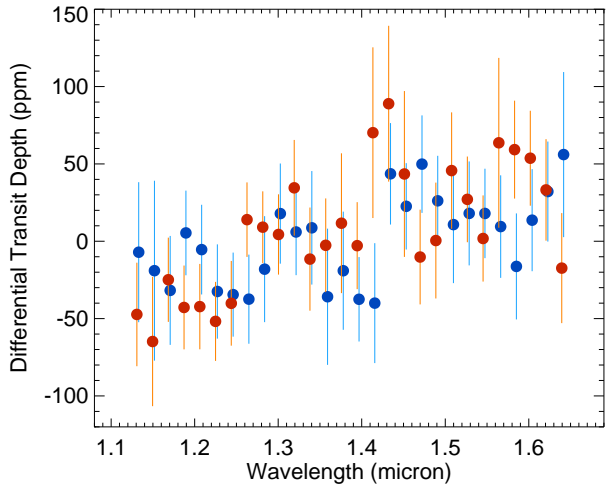}
\centering
\caption{Wavelength-dependent relative transit depths measured for the UT 2013 Dec 19 (blue filled circles) and UT 2014 Jan 7 (red filled circles) visits.  The white-light transit depths have been subtracted from each visit.}
\label{indiv_spec}
\end{figure}

We estimate uncertainties in the best-fit planet star radius ratios for each wavelength element using both a residual permutation method and the covariance matrix from our Levenberg-Marquart minimization, where we set the uncertainties on individual data points in a given wavelength channel equal to the standard deviation of the residuals in that channel after the best-fit model has been subtracted.  We find that the residual permutation errors for both visits are on average higher than the corresponding covariance matrix errors, suggesting the presence of time-correlated noise in our light curves.  However, the residual permutation errors also display a greater variation in size from one wavelength channel to the next, likely as a result of sampling error due to the limited number of data points in each light curve.  We therefore take a conservative approach and select the larger of the two error estimates in each wavelength channel as our final error estimate, which allows us to account for time-correlated noise in our data while avoiding possible under-estimation of uncertainties due to the limited number of measurements available for the residual permutation analysis.

Next we bin the resulting wavelength-dependent radius values by a factor of four to create the transmission spectrum shown in Fig. \ref{indiv_spec}, where we average the estimated radii and corresponding errors in each of the twenty eight bins.  Because we have smoothed our spectra in wavelength space prior to fitting for the transit depths in each individual channel, the transit depths in adjacent wavelength channels are correlated with each other and the errors on the binned transit depths do not decrease as the square root of the number of wavelength channels in the bin.  We estimate the effect of this smoothing on the binned errors by creating a simulated data set with the same time sampling and number of wavelength channels as our original data.  We add noise to these simulated data and fit them in exactly the same way as our original data in order to derive a binned transmission spectrum.  We then repeat this analysis with the same simulated data, but including an additional step where we smooth the simulated ``spectra" using the same Gaussian function that we apply to our actual spectra.  When creating our binned transmission spectrum, we find that we must average the four individual error estimates from the smoothed version of the data and multiply this average by a factor of 1.22 in order to obtain the same binned uncertainties as in the unsmoothed version.  We adopt this same scaling when calculating the errors on the binned transit depth estimates from our real data.  

\begin{deluxetable}{ccccc}
\tablecaption{Wavelength Dependent Transit Depths from Template Fitting Method \label{spec_table1}}
\tablehead{
\colhead{Wavelength} & \colhead{Depth} & \colhead{White Noise Error\tablenotemark{a}} & \colhead{Total Error\tablenotemark{a}} \\
\colhead{\micron} & \colhead{ppm} & \colhead{ppm} & \colhead{ppm} 
}
\startdata
       1.132 &  912 &  20 & 27 \\
       1.151 &  896 & 20 & 34\\
       1.170 &  916 & 20 & 22\\
       1.188 &  917 & 19 & 19\\
       1.207 & 913 & 20 & 20\\
       1.226 &  897 & 19 & 20\\
       1.245 & 898 & 18 & 19\\
       1.264 &  934 & 18 & 18 \\
       1.283 &  948 & 18 & 19 \\
       1.301 &  952 & 19 & 20 \\
       1.320 &  952 & 19 & 21\\
       1.339 & 936 & 18 & 25\\
       1.358 & 933 & 18 & 25\\
       1.377 & 925 & 19 & 29\\
       1.396 & 914 & 19 & 20\\
       1.415 & 922 & 19 & 32\\
       1.433 &  981 & 20 & 27\\
       1.452 &   946 & 19 & 25\\
       1.471 &  956  & 18 & 22 \\
       1.490 & 945 & 18 & 23 \\
       1.509 & 964 & 20 & 27 \\
       1.528 & 965 & 20 & 21 \\
       1.546 & 947 & 19 & 20 \\
       1.565 &  946 & 21 & 28 \\
       1.584 & 963 & 21 & 23 \\
       1.603 & 973 & 20 & 22 \\
       1.622 &  968 & 20 & 23 \\
       1.641 & 953 & 24 & 30 \\
\enddata
\tablenotetext{a}{White noise measurement errors are estimated using the covariance matrix, which implicitly assumes that individual measurement errors are Gaussian distributed and uncorrelated.  The total measurement errors are calculated by comparing the covariance errors to a residual permutation method that better accounts for time-correlated noise and taking the larger of the two in each wavelength channel.  Uncertainty in the limb-darkening models from the stellar effective temperature contributes an additional error of 1 ppm in each bin, which we do not include here.}
\end{deluxetable}

\begin{figure}[ht]
\epsscale{1.2}
\plotone{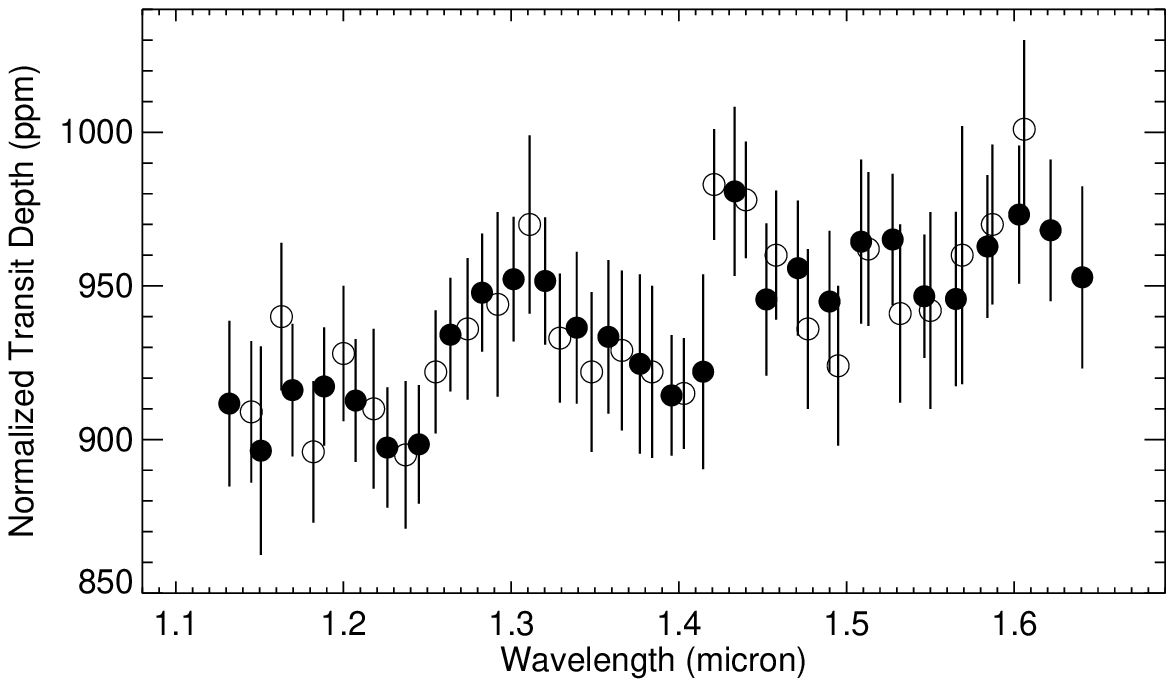}
\centering
\caption{Wavelength-dependent transit depths averaged over the two visits, where the depths are defined as the square of the planet-star radius ratio $R_p/R_\star$ in each band.  Depths derived using the spectral template fitting technique \citep{deming13,knutson14} are shown as filled circles, and depths from the white light residual fitting technique \citep{kreidberg14} are shown as open circles.  No offset has been applied to either data set, demonstrating that the average transit depths are also in good agreement.}
\label{spec_comparison}
\end{figure}

We take the error-weighted average of the two individual transmission spectra to create the combined spectrum shown in Fig. \ref{spec_comparison} and in Table \ref{spec_table1}.  We present two separate estimates for the measurement uncertainties in Table \ref{spec_table1}, including one calculated using the covariance matrix errors alone and the other taking the larger of the two error estimates as previously described.  As an additional consistency check, we also carried out a version of our analysis in which we allowed the light curves derived from the forward and reverse scans to have different planet-star radius ratios.  We found that we obtained consistent radius estimates from both scan directions, indicating that the different behaviors visible in the white-light curves are effectively removed in the differential time series by our template fitting technique.  

\subsection{White Light Residual Fitting Method}

\begin{deluxetable}{ccccc}
\tablecaption{Wavelength Dependent Transit Depths from White Light Residual Method \label{spec_table2}}
\tablehead{
\colhead{Wavelength} & \colhead{Depth} & \colhead{White Error\tablenotemark{a}} &  \colhead{Total Error\tablenotemark{a}} & \colhead{Reduced $\chi^2$} \\
\colhead{\micron} & \colhead{ppm} & \colhead{ppm} 
}
\startdata
1.145 & 909 & 20 & 23 & 0.81 \\
1.163 & 940 & 19 & 24 & 0.77 \\
1.182 & 896 & 19 & 23 & 0.93 \\
1.200 & 928 & 19 & 22 & 0.77 \\
1.218 & 910 & 19 & 26 & 0.82 \\
1.237 & 895 & 18 & 24 & 0.82 \\
1.255 & 922 & 18 & 20 & 0.76 \\
1.274 & 936 & 19 & 23 & 0.88 \\
1.292 & 944 & 18 & 30 & 0.83 \\
1.311 & 970 & 18 & 29 & 1.04 \\
1.329 & 933 & 18 & 21 & 0.87 \\
1.348 & 922 & 18 & 26 & 0.80 \\
1.366 & 929 & 18 & 26 & 0.81 \\
1.384 & 922 & 18 & 28 & 0.83 \\
1.403 & 915 & 18 & 18 & 0.89 \\
1.421 & 983 & 18 & 18 & 0.84 \\
1.440 & 978 & 18 & 19 & 0.85 \\
1.458 & 960 & 18 & 21 & 0.75 \\
1.477 & 936 & 19 & 26 & 0.68 \\
1.495 & 924 & 19 & 26 & 0.89 \\
1.513 & 962 & 19 & 25 & 0.91 \\
1.532 & 941 & 19 & 29 & 0.89 \\
1.550 & 942 & 19 & 32 & 0.97 \\
1.569 & 960 & 20 & 42 & 0.95 \\
1.587 & 970 & 20 & 26 & 0.87 \\
1.606 & 1001 & 20 & 29 & 0.87 \\
\enddata
\tablenotetext{a}{White noise measurement errors are estimated using a Markov Chain Monte Carlo analysis, which implicitly assumes that individual measurement errors are Gaussian distributed and uncorrelated.  Total measurement errors are calculated by comparing the MCMC errors to a residual permutation method that better accounts for time-correlated noise and taking the larger of the two in each wavelength bin.}
\end{deluxetable}

\begin{figure}[ht]
\epsscale{1.2}
\plotone{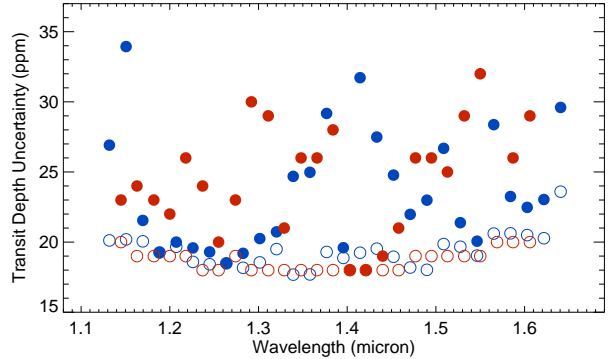}
\centering
\caption{Comparison of uncertainties on the reported transit depths under the assumption of either white, Gaussian noise (open circles) or allowing for time-correlated noise using the residual permutation method for estimating uncertainties (filled circles).  Errors derived using the spectral template fitting technique \citep{deming13,knutson14} are shown as blue circles, and depths from the white light residual fitting technique \citep{kreidberg14} are shown as red circles.}
\label{error_comparison}
\end{figure}

We also obtained an independent estimate of the transmission spectrum following the approach of \citet{kreidberg14} and \citet{stevenson14}.  Beginning with the raw images, this analysis used an entirely different set of codes than those described in the previous sections.  In this version of the analysis we treated each up-the-ramp sample in the ima.fits images as an independent subexposure.  For each subexposure, we interpolated all rows to a common wavelength scale to account for the changing dispersion solution with spatial position on the detector.  We estimated the background by making conservative masks around the stellar spectra and taking the median of the unmasked pixels.  We subtracted the background and optimally extracted the spectra.  To combine the data from individual subexposures, we summed the spectra by column.  The final step in the reduction process is to correct for drift of the spectra in the dispersion direction over the course of a visit.  We used the first exposure from the first visit as a template and shifted all subsequent spectra to the template wavelength scale.  The spectra shifted by a total of 0.3 pixels over the five orbits contained in our observations, which is larger than the approximately 0.01 pixel drift observed in previous scanning mode observations of GJ 1214b \citep{kreidberg14}.  This increased drift may be related to the longer scan length and faster scan rate utilized for these observations as compared to GJ 1214b.  

We binned the spectra in four-pixel-wide channels, yielding 26 spectrophotometric light curves between 1.15 and 1.61~\micron.  The light curves show orbit-long ramp-like systematics that are characteristic of WFC3 data.  We correct for these systematics using the divide-white technique, which assumes that the observed effects have the same shape across all wavelengths.  We fit each spectroscopic light curve with a transit model multiplied by a scaled vector of systematics from the best-fit white light curve and a linear function of time.   We fix $i$ and $a/R_\star$ to the values reported in Table \ref{transit_fit}.  The fit to each channel has six free parameters: one transit depth, four scaling factors for the systematics (one for each visit and scan direction), and the linear slope.  As before, we calculate the four-parameter nonlinear limb-darkening coefficients using a \texttt{PHOENIX} stellar atmosphere model where we take the flux-weighted average of the theoretical stellar intensity profile within each photometric bandpass.  We set the uncertainties on individual points equal to the sum of the the photon noise and read noise in quadrature.  We list the best-fit values and corresponding errors in Table \ref{spec_table2}.  

We report uncertainties on the transit depths corresponding to $1\sigma$ confidence intervals from either a Markov chain Monte Carlo (MCMC) fit, which implicitly assumes white (Gaussian and uncorrelated) noise, or a residual permutation analysis that better accounts for any time-correlated noise present in the data.  We take the larger of the two errors in each wavelength bin as our final uncertainties and provide the MCMC only errors separately in Table \ref{spec_table2} for comparison.  Our residual permutation errors are on average 30\% larger than those obtained with MCMC.  This suggests that there is some time-correlated noise in the light curves, which is most likely the result of imperfect corrections for visit- and orbit-long systematic trends in the data.

\section{Discussion and Conclusions}\label{discussion}

\begin{figure}[ht]
\epsscale{1.2}
\plotone{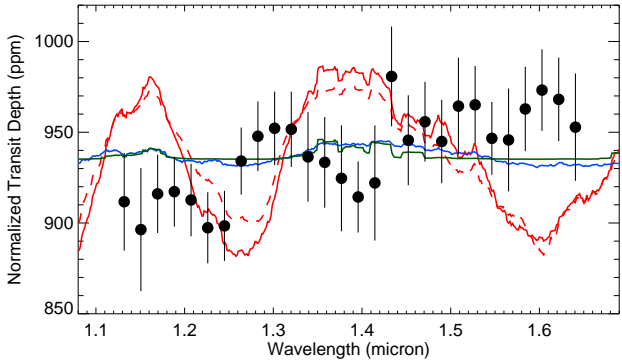}
\centering
\caption{Wavelength-dependent transit depths averaged over the two visits (black filled circles).  Four different atmosphere models are shown for comparison:  a solar metallicity model (red solid line), a $50\times$ solar metallicity model (red dashed line), a pure water model (solid blue line), and a solar metallicity model with an opaque cloud deck at 1 mbar (solid green line).  The average depth for each model has been normalized to match the average measured transit depth in this plot.}
\label{avg_spec}
\end{figure}

We used the the WFC3 instrument on the \textit{Hubble Space Telescope} to observe transits of the super-Earth HD 97658b at wavelengths between $1.1-1.7$~\micron.  Our white-light transit fits produce consistent estimates of the transit times and planet star radius ratios between the two visits.  Our errors for these parameters are dominated by systematic noise, as illustrated in Fig. $1-3$.  We combine our measured transit times with previously published values from \textit{MOST} and \textit{Spitzer} and show that they are consistent with a linear ephemeris.  We also derive an improved estimate for the planet's orbital period with an uncertainty that is a factor of thirty smaller than in \citet{vangrootel14}.  Our average radius ratio is $2.5\sigma$ larger than the \textit{Spitzer} 4.5~\micron~measurement from \citet{vangrootel14}, and is in good agreement with the \textit{MOST} visible-light value from \citet{dragomir13}.  It is unlikely that this offset is due to stellar activity as this star has a \logr~value between $-4.95$ and $-5.00$ \citep{howard11} and varied by less than $0.2\%$ in brightness over a single 1.5 day visible-light observation by \textit{MOST} \citep{dragomir13}.

We obtain nearly identical versions of the transmission spectrum using the spectral template fitting technique \citep{deming13,knutson14} and the white light residual method \citep{kreidberg14}, as shown in Fig. \ref{spec_comparison}.  Our median uncertainties on the differential wavelength-dependent transit depth are 23 ppm from the spectral template fitting and 26 ppm from the white-light residual method (see Fig. \ref{error_comparison}), making these observations the most precise measurement of a planetary transmission spectrum that we are aware of with WFC3 to date.  Both versions of our spectrum display a discontinuity around 1.42~\micron, which may be related to a slight upward slope visible towards redder wavelengths.  These features could be instrumental in nature, as the columns that make up the wavelength bins closest to the discontinuity are located on a boundary between quadrants on the WFC3 array.  The statistical significance of these features is marginal, and we therefore do not consider them in our subsequent comparison to atmosphere models for this planet.

In Fig. \ref{avg_spec} we compare our measured transmission spectrum from the spectral template fitting technique to several representative atmosphere models for HD 97658b, which are calculated following \citet{kempton12}.  The effective temperatures ($720-730$~K, depending on the model) and corresponding pressure-temperature profiles of these models are calculated assuming full redistribution of energy to the planet's night side and an albedo which varies according to composition.  We consider both cloud-free models with solar and $50\times$ solar metallicities, as well as a pure water model and a solar metallicity model with a high cloud deck located at one mbar.  We calculate the significance with which our data can rule out a given model using the equation from \citet{gregory05}:

\begin{equation}\label{eq3}
Significance = \frac{\chi^2-\nu}{\sqrt{2\nu}}
\end{equation}
where $\chi^2$ is calculated by comparing our averaged transmission spectrum to each model and $\nu$ is the number of degrees of freedom in the fit (27 in our case, as there are 28 points and we normalize the models to match the average measured transit depth of our data).  This metric assumes that our measurement errors are Gaussian and uncorrelated from one wavelength bin to the next; although this is almost certainly untrue at some level, it represents a reasonable starting point for comparing different models.  Following this approach we find that our measured transmission spectrum is inconsistent with the solar and $50\times$ solar cloud-free models at the $10\sigma$ and $9\sigma$ levels, respectively.  It is equally well described by the water-dominated ($0.6\sigma$) model and the solar metallicity model with optically thick clouds at a pressure of one mbar ($0.6\sigma$), as well as a flat line at the average transit depth across the band ($0.4\sigma$).  We find that a solar metallicity model with clouds at 1 mbar is also consistent with the data ($1.8\sigma$), indicating that the clouds could be located slightly deeper in the atmosphere.  We note that there are any number of high metallicity atmosphere models that could provide a fit comparable to that of the pure water model; all our data appear to require is either a relatively metal-rich atmosphere with a correspondingly small scale height, or the presence of a high cloud deck that obscures the expected water absorption feature in a hydrogen-dominated atmosphere.  We constrain the maximum hydrogen content of the atmosphere in the first scenario by considering a series of cloud-free models with varying number fractions of molecular hydrogen and water, and find that in this scenario the atmosphere has to be at least $20\%$ water by number in order to be consistent with our data at the $3\sigma$ level.  We list the $\chi^2$ values and level of disagreement for all models in Table \ref{atm_comparison}.

 \begin{deluxetable}{lccc}
\tablecaption{Atmosphere Model Comparison \label{atm_comparison}}
\tablehead{
\colhead{Model} & \colhead{$\chi^2$} & \colhead{Level of Disagreement\tablenotemark{a}}
}
\startdata
Flat line & 29.9 & 0.4$\sigma$ \\
Solar metallicity & 102.9 & 10.3$\sigma$ \\
50x solar & 92.2 & 8.9$\sigma$ \\
Solar with cloud deck at 1 mbar  & 31.2 & 0.6$\sigma$ \\
Solar with cloud deck at 10 mbar & 40.4 & 1.8$\sigma$ \\
10\% H$_2$O & 53.6 & 3.6$\sigma$ \\
20\% H$_2$O & 42.2 & 2.1$\sigma$ \\
30\% H$_2$O & 37.5 & 1.4$\sigma$ \\
40\% H$_2$O & 35.1 & 1.1$\sigma$ \\
100\% H$_2$O & 31.1 & 0.6$\sigma$ \\
\enddata
\tablenotetext{a}{This is the significance with which we can rule out a given model, calculated according to Eq. \ref{eq3}.  Significance levels less than $3\sigma$ mean that the data are consistent with that model within the reported errors.}
\end{deluxetable}

The conclusion that HD 97658b's transmission spectrum appears to be flat at the precision of our data places it in the same category as both the super-Earth GJ 1214b \citep{kreidberg14} and the Neptune-mass GJ 436b \citep{knutson14}.  As with these two planets, a more precise measurement of HD 97658b's transmission spectrum will eventually allow us to distinguish between high clouds and a cloud-free, metal-rich atmosphere.  Our constraints on the atmospheric scale height in the cloud-free scenario are relatively weak compared to those obtained for GJ 1214b and GJ 436b, despite the fact that we achieve smaller errors (20 ppm vs 30 ppm) in our measurement of the differential transmission spectrum.  This is primarily because HD 97658b has a smaller planet star radius ratio than either of these systems, and the predicted amplitude of the transmission spectrum is correspondingly small.  Fortunately, it also orbits a brighter star than either GJ 1214b or GJ 436b, making it possible to achieve high precision transit measurements with relatively few observations.  Unfortunately, this makes ground-based observations particularly challenging as the nearest comparison star with a comparable brightness is located more than 40\arcmin~away.  For space telescopes such as \textit{Hubble} and \textit{Spitzer}, achieving the precision required to study this planet in detail will mean pushing the systematic noise floor to unprecedentedly low levels.  There is every reason to believe that this level of performance should be achievable, and given the unique nature of this planet it is likely that this will be put to the test in the near future.

\acknowledgments

We are grateful to STScI Director Matt Mountain and the Director's Discretionary Time program for making it possible to obtain these observations prior to the next regular \textit{HST} proposal cycle.  We would also like to thank Beth Perriello and the rest of the \textit{HST} scheduling team, as they worked overtime in the days leading up to the Christmas holiday in order to make sure that we obtained the best possible data from this program.  We also wish to acknowledge Valerie Van Grootel, who was kind enough to share an advance copy of her paper prior to publication.  H. K. acknowledges support from NASA through grant GO-13501.  L.K. received funding for this work from the National Science Foundation in the form of a Graduate Research Fellowship. J.L.B. acknowledges support from the Alfred P. Sloan Foundation and NASA through grants NNX13AJ16G, GO-13021, and GO-13467.  D. H. has been supported through the European Research Council advanced grant PEPS awarded to Gilles Chabrier under the European CommunityÕs Seventh Framework Program (FP7/2007-2013 Grant Agreement no. 247060).

%{\it Facilities:} \facility{Keck:I (HIRES)}, \facility{Keck:II (NIRC2)}

\end{document}